\documentclass[forecasting,article,accept,pdftex,oneauthor]{Definitions/mdpi} 

\firstpage{1} 
\pubvolume{1}
\issuenum{1}
\articlenumber{0}
\pubyear{2023}
\copyrightyear{2023}
\externaleditor{{Academic Editor: } 
}
\datereceived{ } 
\daterevised{ } 
\dateaccepted{ } 
\datepublished{ } 
\hreflink{https://doi.org/} 

\Title{Comparative Analysis of Machine Learning, Hybrid, and Deep Learning Forecasting Models: Evidence from European Financial Markets and Bitcoins}

\TitleCitation{Comparative Analysis of Machine Learning, Hybrid, and Deep Learning Forecasting Models: Evidence from European Financial Markets and Bitcoins}

\Author{{Apostolos Ampountolas} $^{1}$*$^{2}$\orcidA{}}

\AuthorNames{Apostolos Ampountolas}

\AuthorCitation{Ampountolas, Apostolos}

\address{%
$^{1}$ \quad School of Hospitality Administration, Boston University, Boston, MA 02215, USA; aampount@bu.edu\\
$^{2}$ \quad Department of Mathematics,  College of Engineering, Design and Physical Sciences, \mbox{Brunel University London}, {Uxbridge} UB8 3PH, UK; apostolos.ampountolas@brunel.ac.uk}

\corres{Correspondence: aampount@bu.edu}

\abstract{This study analyzes the transmission of market uncertainty on key European financial markets and the cryptocurrency market over an extended period, encompassing the pre-, during, and post-pandemic periods. Daily financial market indices and price observations are used to assess the forecasting models. We compare statistical, machine learning, and deep learning forecasting models to evaluate the financial markets, such as the ARIMA, hybrid ETS-ANN, and \textit{k}NN predictive models. The study results indicate that predicting financial market fluctuations is challenging, and the accuracy levels are generally low in several instances. ARIMA and hybrid ETS-ANN models perform better over extended periods compared to the \textit{k}NN model, with ARIMA being the best-performing model in 2018--2021 and the hybrid ETS-ANN model being the best-performing model in most of the other subperiods. Still, the \textit{k}NN model outperforms the others in several periods, depending on the observed accuracy measure. Researchers have advocated using parametric and non-parametric modeling combinations to generate better results. In this study, the results suggest that the hybrid ETS-ANN model is the best-performing model despite its moderate level of accuracy. Thus, the hybrid ETS-ANN model is a promising financial time series forecasting approach. The findings offer financial analysts an additional source that can provide valuable insights for investment decisions.}


\begin{document}
\section{Introduction}
The worldwide spread of the COVID-19 pandemic has resulted in a significant economic impact that has intensified market risk aversion to a degree that has not been observed since the global financial crisis. During~this period, ~financial markets worldwide have experienced a significant decline since January 2020. This disruption has resulted in highly high fluctuations and unpredictability. Despite the ongoing pandemic and its indeterminate long-term effects, the~U.S. government has recently declared an end to the pandemic \citep{NYT2022COVID}, likely due to the transition to the endemic stage of the COVID-19 outbreak, characterized by the virus's widespread presence but significantly reduced fatalities compared to 2020 \citep{fisher2021impact}. According to a study conducted by~\cite{ashraf2020economic}, there is an adverse correlation between the number of confirmed COVID-19 cases and the U.S. stock market returns. The~study further found that stock markets responded reasonably quickly to the increase in confirmed cases rather than deaths. Therefore, the~impact of the crisis on financial markets and the broader economic environment, including consumer behavior and intentions, has been a topic of concern among many scholars \citep{ashraf2020economic, watson2021will, MAZUR2021101690, goldstein2021covid, di2022covid}. From~this standpoint, asset management has become crucial for organizations seeking to optimize their asset utilization and maximize returns in challenging conditions and volatile~markets.

The ongoing pandemic and its consequential impact on financial markets have already elicited significant scholarly attention. In~the near future, this outbreak is expected to serve as a standard for comparison, exceeding the level of interest observed during previous crises of similar nature. Developing models that yield accurate stock market fluctuation predictions is fundamental to improving the informational base, meeting the investor's goals and return expectations, and~accommodating risk tolerance over a specified investment period. This is because asset allocation is a strategic approach that seeks to attain an equilibrium between risk and return by adjusting the proportion of each asset in a portfolio. Despite this exercise's significance for risk management predictions, only a few prior studies (see~\cite{aslam2020evidence, khattak2021predicting, su2022covid19, lachaab2023machine}) have focused on the European stock market and attempted to estimate the key financial markets' volatility. At~the same time, the~financial markets literature on forecasting cryptocurrencies, U.S. stock, and~commodity market volatility is more mature; refer, for~example, to~recent academic research by~\cite{topcu2020impact, azimli2020impact, MAZUR2021101690, goodell2021co, akhtaruzzaman2022systemic, uddin2022stock, ampountolas2023effect}), which has made predictions regarding significant incidents, such as COVID-19 and its economic consequences, while also examining the impact of past~pandemics.

Therefore, this study aims to address the discrepancy in the literature on forecasting market fluctuations by examining the European stock market indices. The~availability of financial market information has increased, strengthening the correlation between market distribution and other factors. Over~the past decades, a~significant body of research has grown on forecasting time series using various linear, nonlinear, and~hybrid linear-nonlinear models. In~this study, we employed the application of traditional statistical models, such as the autoregressive integrated moving average (ARIMA) \citep{hyndman2008automatic}, as~well as machine learning (ML), deep learning (DL), such as the long short-term memory (LSTM) and hybrid forecasting models, i.e.,~the ETS-ANN model \citep{panigrahi2017hybrid} to evaluate the volatility of the European stock markets. The~European stock market encompasses a wide range of stock exchanges, including major ones, such as the London Stock Exchange (LTSE), Euronext (N100), Deutsche Börse (GDAXI), the~French stock market (FCHI), and~the Swiss Stock Exchange (SSMI). 

This study contributes twofold to the expanding body of literature in the field. First, it examines the distribution of fluctuations between the European stock market and cryptocurrency in three subperiods: before, during, and~ongoing the COVID-19 outbreak. We chose to examine the European stock market because the COVID-19 pandemic has, from~one side, severely impacted each European country. Still, on~the other side, each government has imposed diverse restrictions within various time frames. For~the financial markets, the~COVID-19 pandemic constitutes a systematic risk factor. In~this context, government-imposed restrictions on spreading the virus could not prevent the negative consequences of the financial stock markets' trajectory \citep{ashraf2020economic}. However, observing the cryptocurrency market, we picture a different likelihood (see, Figure~\ref{price_returns}). Hence, examining the correlation between the European stock market and the cryptocurrency markets during this period is critical. This analysis of the cross-market relationships offers valuable information for financial market players concerning financial risk management. Bitcoin is the most well-known cryptocurrency with the highest capitalization and it serves as a benchmark for investors and academics. Hence, any disruptions in this market may also have ripple effects on the global financial market. Second, exploring various forecasting models, such as statistical, ML, DL, and~hybrid models, we examine the dynamics of market distribution to provide an understanding of the market predictability among the European stock markets and Bitcoin during the COVID-19 pandemic through an analysis of daily~data.

Our results show that predicting BTC-USD distributions is challenging, and~the accuracy levels are generally low, especially in 2018 and 2019. GDAXI is more predictive than BTC-USD, and~the ETS-ANN and \textit{k}NN models perform both in different subperiods. The~ARIMA and ETS-ANN models perform better than the \textit{k}NN model for predicting the FTSE index's performance, but~their performances were generally weak in 2018--2021 and 2020. ARIMA and ETS-ANN models perform better over extended periods than the \textit{k}NN model, with~ARIMA being the best-performing model in 2018--2021 and the hybrid ETS-ANN model in 2020. The~hybrid ETS-ANN model shows significant potential for forecasting financial time series in this~context.

The subsequent sections of the document are organized as follows: Section~\ref{sec2} provides a short review of the literature on financial market forecasts through machine learning techniques. Section~\ref{sec3} discusses the data collection and modeling methodologies. Section~\ref{sec4} compares the different models' accuracy measures and outcomes. The~article concludes with the fifth section, including future research~suggestions.

\section{Research~Background}\label{sec2}
Numerous unforeseen and rare events, called black swan events, significantly impact the stock market. A~growing body of scholarly literature has examined the effects of the COVID-19 pandemic on the stock market, mainly concentrating on the US market, and~observed how stock and cryptocurrency markets reacted to various global events; for example, {see} 
 (\cite{leippold2022machine, CINER2021101705, ampountolas2023effect, di2022covid, uddin2022stock, ijfs10030051}). 
 \cite{azimli2020impact} conducted a study to examine the influence of COVID-19 on the relationship between return and risk through a quantile regression methodology. 
~\cite{al2020death} reported that the confirmed COVID-19 cases' daily growth rate and the total number of associated deaths significantly impacted the stock returns in the Chinese stock markets. Furthermore, there is a contention that the scale of the impact of disease epidemics has been comparatively less substantial than that of the COVID-19 pandemic. ~\cite{liu2020covid} performed an event study to assess the instantaneous effects of the COVID-19 pandemic on 21 stock indices in countries significantly affected by the outbreak. The~results indicate that the stock market's reaction was swift, leading to an immediate decrease, with~Asian stock markets demonstrating higher negative abnormal returns than other countries. Another study used the wavelet approach to investigate the multiscale co-movement between Bitcoin and COVID-19 deaths \citep{goodell2021co}. 

This study centers on the pandemic period and the corresponding impact on key European financial markets compared to cryptocurrency, specifically Bitcoin. In~recent literature, one of the first studies~\cite{aslam2020evidence} assessed the effects of the COVID-19 pandemic on eight European stock markets. They utilized multifractal detrended fluctuation analysis (MFDFA) to calculate the Hurst exponents. Their findings validate the presence of multifractal characteristics in the European stock markets amidst the COVID-19 pandemic. As~such, the~efficiency levels of these markets differ based on their multifractal properties. For~example, the~efficiency of the Spanish stock market is comparatively higher than that of Austria, which exhibits a relatively lower efficiency. Belgium, Italy, and~Germany are positioned within the intermediate range. In~a recent study, \cite{lachaab2023machine} examined the accuracy of machine learning and deep learning techniques in forecasting the FCHI 40 index during the COVID-19 outbreak. The~study aimed to determine whether the index and individual prices would maintain the steady growth they attained at the start of the vaccination administration. The~authors used \textit{k}NN and LSTM and assessed their performances compared to the ARIMA time series model. Their findings identified that the \textit{k}NN technique outperformed LSTM and ARIMA. ~\cite{su2022covid19} examined the adverse effects of the COVID-19 pandemic on the return dynamics of 10 European stock markets before and after the onset of the COVID-19 pandemic. Their results were unexpectedly positive, as~they observed a swift and unparalleled recovery in the European stock market, resulting in notable yield gains after the COVID-19 outbreak. In~another study, ~\cite{khattak2021predicting} employed the least absolute shrinkage and selection operator (LASSO) to examine 21 potential internal and external shocks to the European financial market during the COVID-19 outbreak. They found Germany and France to be the most critical determinants of the European~market.

\section{Data and~Methodology}\label{sec3}
\unskip
\subsection{Data~Analysis}
This study evaluated daily prices for the BTC-USD ({{data are publicly available at} 
 {\url{https://www.cryptodatadownload.com/data/}} (accessed on 22 January 2023) for the cryptocurrency in the EDT time zone}) and the European financial markets FCHI, GDAXI, FTSE, SSMI, and~N100 indices. The~daily price dataset was obtained from Yahoo {Finance } 
(\url{finance.yahoo.com}, (accessed on 22 January 2023)) for the period that ranges from {1 January 2018} to~{31 December 2021}. To~forecast the index fluctuations and evaluate the various models, we divided the sample dataset into three subperiods: (i) the pre-COVID period that ranges from {1 January 2018} 
 to~{31 December 2019}; (ii) the main COVID period that ranges from {1 January 2020} to~{31 December 2020}; and (iii) the ongoing COVID-19 period ranges from {1 January 2021} to~{31 December 2021}. We split the dataset into two segments for each subperiod: a training set and a test set to assess the model's forecasting performance, following a ratio of 80:20. To~evaluate the study's models, stationary data are required; hence, we computed the log returns and ran the augmented Dickey--Fuller (ADF) tests to test the unit root null hypothesis. Finally, the~Jarque--Bera test will define whether or not the financial asset dataset has skewness and kurtosis corresponding to a normal~distribution.

\begin{figure}[H]
\begin{adjustwidth}{-\extralength}{0cm}
\centering 
	\includegraphics[width=1.05\textwidth]{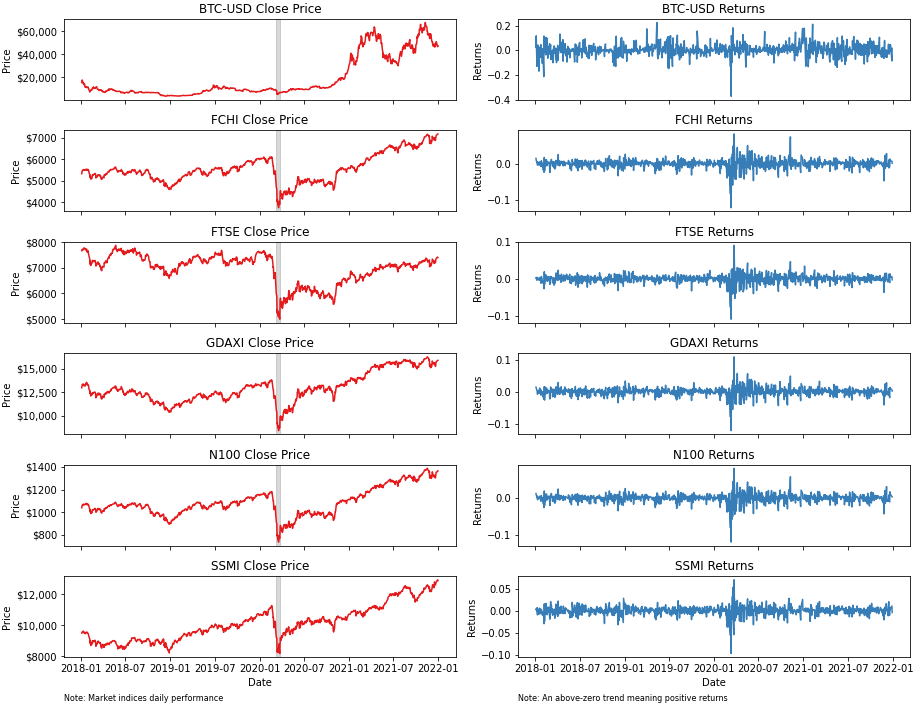}
\end{adjustwidth}
\caption{{Financial market price} 
 and returns trajectory---period 2018--2021.} \label{price_returns}	
\end{figure}

Figure~\ref{price_returns} illustrates the daily close price and returns the trajectory of the financial market indices dataset. We initially noticed both substantial increasing and downward tendencies. Thus, we observed a noticeable sudden shift when the World Health Organization (WHO) denoted COVID-19 as a pandemic on {11 March 2020}, in~all financial market indices, but~significantly less for the BTC market. Consequently, the~fat tails and volatility clustering are more significant during the pandemic than during the remaining period, which follows a similar pattern. In~addition, we observed that in most of the examined indices, the~stock prices returned to their pre-crash levels following a brief period of market disruption during the trading days of the financial market crash. However, in~most instances, the~increase is developing moderately, except~for the GDAXI index, which reacted positively and swiftly following a few trading days. After~a bear market, the~market has shown evidence of a sharp recovery. As~a result, we experienced a momentum shift in market conditions as the market returned to upward movement after the previous change in market~movements.

Table~\ref{tab:descriptive} presents the summary statistics of the financial asset returns for the entire period, divided into four subperiods, i.e.,~pre-COVID-19, during~the COVID-19 period, and~finally, the~ongoing COVID-19 period in the year 2021. We encounter positive mean returns for the entire financial asset indices before the pandemic and positive ones during the pandemic, except~for the FCHI (French market) and the FTSE (London Stock Exchange). Interestingly, the~mean returns of the BTC were higher during the COVID-19 period than in any other period. At~the same time, it had higher negative returns in 2018. All markets show increased volatility, following a similar pattern throughout the period. The~BTC market has the highest volatility, followed by the GDAXI and the FCHI markets for different periods. We also reported the Sharpe ratio, which is widely used among investors to evaluate investment performance. The~Sharpe ratio reports, except~for BTC, negative values for the remaining financial asset returns throughout the study period. The~unit root ADF (augmented Dickey--Fuller) stationarity test suggests that the financial asset return series are stationary throughout all subperiods. Finally, the~Jarque--Bera test statistic indicates how closely the returns correspond to a distribution that is normally distributed. The~results of the Jarque--Bera test confirm evidence of non-Gaussian distributions, and~we observed high positive values during COVID-19 in~2020.

\begin{table}[H]
\caption{Descriptive statistics for the financial market~returns.}
\label{tab:descriptive}
\setlength{\tabcolsep}{6pt}
\renewcommand{\arraystretch}{1.1}
\resizebox{\textwidth}{!}{%
\begin{tabular}{lllllllll}
\toprule
\textbf{Index}      & \textbf{Mean}      & \textbf{Std Dev}   & \textbf{Min}      & \textbf{Max}    & \textbf{SR}     & \textbf{SE}     &\textbf{ ADF Stat }& \textbf{JB-Stat } \\ \midrule
\multicolumn{5}{l}{Panel A: ({1 January 2018--31 December 2018}
)}       &        &        &          &          \\
BTC-USD    & $-$0.0030   & 0.0422    & $-$0.1685  & 0.1322 & $-$0.4289 & 0.0022 & $-$6.75 *** & 50.58 *** \\
FCHI        & $-$0.0003   & 0.0073    & $-$0.0332  & 0.0262 & $-$0.8446 & 0.0004 & $-$18.51 ***   & 99.73 ***    \\
FTSE       & $-$0.0003   & 0.0067    & $-$0.0315  & 0.0235 & $-$0.9122 & 0.0004 & $-$19.81 ***   & 124.61 ***   \\
GDAXI      & $-$0.0005   & 0.0081    & $-$0.0348  & 0.0290 & $-$0.8578 & 0.0004 & $-$20.11 ***   & 57.25 ***    \\
N100   & $-$0.0003   & 0.0069    & $-$0.0342  & 0.0252 & $-$0.9059 & 0.0004 & $-$18.75 ***   & 172.41 ***   \\
SSMI        & $-$0.0003   & 0.0075    & $-$0.0313  & 0.0285 & $-$0.8031 & 0.0004 & $-$20.21 ***   & 138.34 ***   \\ \midrule
\multicolumn{5}{l}{Panel B: ({1 January 2019--31 December 2019})}        &        &        &          &          \\
BTC-USD    & 0.0024    & 0.0356    & $-$0.1409  & 0.1736 & 0.1426 & 0.0019 & $-$19.63 ***   & 343.33 ***   \\
FCHI        & 0.0007    & 0.0070    & $-$0.0357  & 0.0272 & $-$0.2569 & 0.0004 & $-$19.23 ***   & 346.17 ***   \\
FTSE       & 0.0003    & 0.0062    & $-$0.0323  & 0.0225 & $-$0.5346 & 0.0003 & $-$18.52 ***   & 286.89 ***   \\
GDAXI      & 0.0006    & 0.0073    & $-$0.0311  & 0.0337 & $-$0.2735 & 0.0004 & $-$19.44 ***   & 253.62 ***   \\
N100   & 0.0006    & 0.0066    & $-$0.0328  & 0.0266 & $-$0.2951 & 0.0003 & $-$19.42 ***   & 323.99 ***   \\
SSMI        & 0.0006    & 0.0055    & $-$0.0208  & 0.0228 & $-$0.3609 & 0.0003 & $-$18.11 ***   & 97.09 ***    \\  \midrule
\multicolumn{5}{l}{Panel C: ({1 January 2020--31 December 2020})}        &        &        &          &          \\
BTC-USD    & 0.0046    & 0.0377    & $-$0.3717  & 0.1819 & 0.3790 & 0.0020 & $-$8.88 ***    & 12,224.20 *** \\
FCHI        & $-$0.0001   & 0.0171    & $-$0.1228  & 0.0839 & $-$0.2841 & 0.0009 & $-$5.35 ***    & 2144.66 ***  \\
FTSE       & $-$0.0003   & 0.0153    & $-$0.1087  & 0.0905 & $-$0.3855 & 0.0008 & $-$5.36 ***    & 2139.88 ***  \\
GDAXI      & 0.0002    & 0.0173    & $-$0.1224  & 0.1098 & $-$0.2139 & 0.0009 & $-$4.80 ***    & 2507.75 ***  \\
N100   & 0.0000    & 0.0159    & $-$0.1197  & 0.0818 & $-$0.2837 & 0.0008 & $-$5.33 ***    & 2671.55 ***  \\
SSMI        & 0.0001    & 0.0125    & $-$0.0964  & 0.0702 & $-$0.3435 & 0.0007 & $-$5.14 ***    & 3246.57 ***  \\  \midrule
\multicolumn{5}{l}{Panel D: ({1 January 2021--31 December 2021})}       &        &        &          &          \\
BTC-USD    & 0.0022    & 0.0421    & $-$0.1377  & 0.1875 & 0.4751 & 0.0022 & $-$20.06 ***   & 35.57 ***    \\
FCHI        & 0.0007    & 0.0074    & $-$0.0475  & 0.0291 & $-$1.2687 & 0.0004 & $-$21.08 ***   & 652.70 ***   \\
FTSE       & 0.0004    & 0.0067    & $-$0.0364  & 0.0347 & $-$0.8552 & 0.0004 & $-$10.36 ***   & 535.81 ***   \\
GDAXI      & 0.0004    & 0.0076    & $-$0.0415  & 0.0331 & $-$0.8858 & 0.0004 & $-$22.11 ***   & 351.68 ***   \\
N100   & 0.0006    & 0.0072    & $-$0.0428  & 0.0318 & $-$1.1399 & 0.0004 & $-$20.95 ***   & 428.27 ***   \\
SSMI        & 0.0005    & 0.0056    & $-$0.0238  & 0.0210 & $-$1.1433 & 0.0003 & $-$5.38 ***    & 125.07 ***  \\ \bottomrule
\end{tabular}%
}
\noindent{\footnotesize{Note: *** Significant at the 0.001 level. SR defines the Sharpe ratio; SE represents the standard error; ADF represents the augmented Dickey--Fuller test statistics; and JB defines the Jarque--Bera normality test statistics.}}
\end{table}

Table~\ref{tab:correlation} presents the correlation coefficients of the financial asset's daily returns. We observe a modest to high linearity between the European market financial asset indices. It shows that the financial asset indices correlate highly positively with significant correlation coefficients, i.e.,~the correlation between Euronext and the French market FCHI is strongly positive at 0.989, while the correlation between Euronext and the GDAXI index is 0.948. Conversely, BTC and the European market indices have a much smaller linear relationship. The~data indicate a weak positive correlation between Bitcoin and the French CAC 40 index (FCHI), with~a correlation of 0.276. However, Bitcoin has a slightly similar correlation of 0.284 with the EURO STOXX 100 index (N100), indicating a slight tendency for them to move in the same direction. The~correlation between Bitcoin and the German DAX index (GDAXI) is also weak, with~a correlation of 0.274. In~addition, Bitcoin has a weak significant linear relationship with the FTSE, with~a correlation of 0.263. Lastly, the~correlation between Bitcoin and the Swiss Market Index (SSMI) is also weak, with~a correlation of 0.274. Overall, the~correlations in this table suggest very weak linear relationships between Bitcoin and the European financial indices; however, there is a highly positive correlation between them. This is somewhat unique, as~one would expect a higher correlation due to similar exposure to the broader market. Still, it is evident from Figure~\ref{price_returns} that the price trajectory during the study period was~inverse.

\begin{table}[H]
\caption{Financial market returns correlation---period 2018--2021.}
\label{tab:correlation}
\setlength{\tabcolsep}{17pt}
\renewcommand{\arraystretch}{1.0}
\resizebox{\textwidth}{!}{%
\begin{tabular}{@{}lrrrrrr@{}}
\toprule
        & \multicolumn{1}{c}{\textbf{BTC-USD}} & \multicolumn{1}{c}{\textbf{FCHI}} & \multicolumn{1}{c}{\textbf{FTSE}} & \multicolumn{1}{c}{\textbf{GDAXI}} & \multicolumn{1}{c}{\textbf{N100}} & \multicolumn{1}{c}{\textbf{SSMI}} \\ \midrule
BTC-USD & 1                           &                          &                          &                           &                          &                          \\
FCHI    & 0.276                       & 1                        &                          &                           &                          &                          \\
FTSE    & 0.263                       & 0.900                    & 1                        &                           &                          &                          \\
GDAXI   & 0.274                       & 0.944                    & 0.871                    & 1                         &                          &                          \\
N100    & 0.284                       & 0.989                    & 0.913                    & 0.948                     & 1                        &                          \\
SSMI    & 0.274                       & 0.830                    & 0.816                    & 0.814                     & 0.851                    & 1                        \\ \bottomrule
\end{tabular}%
}
\noindent{\footnotesize{Note: Significant at the 0.001 level. Bitcoin (BTC-USD), German DAX (GDAXI), Financial Times Stock Exchange (FTSE), Euronext (N100), French Stock Market (FCHI), Swiss SSMI (SSMI).}}
\end{table}
\unskip

\subsection{Methodology}

\subsubsection{Data~Pre-Processing}
Before using the dataset for the forecasting models, the~original series undergoes two pre-processing techniques: seasonal adjustments and scaling. According to existing literature, it has been observed that neural networks (NNs) often encounter difficulties in effectively capturing the seasonal elements of time series data \citep{barker2020machine}. As~per the methodology, we employed the ADF stationarity test, which suggests the asset return series are~stationary.

Secondly, scaling data is a widely used practice in running neural network forecasting models \citep{shanker1996effect, shankar2016neural}. It involves the normalization of data, typically through utilizing Z-score normalization or min--max normalization as standard preprocessing techniques. The~normalization process has been found to enhance the capacity of models to detect comparable patterns in data that exhibit varying scales \citep{barker2020machine}. Additionally, it has been observed that normalization facilitates the convergence of gradient descent algorithms. As~such, the~scaling application is implemented on the seasonally adjusted series, contingent upon identifying the original series as seasonal. The~simple min--max transformation scales the series within the target range of [0, 1].

\subsubsection{Autoregressive Integrated Moving Average (ARIMA)}
The autoregressive integrated moving average (ARIMA) model is a statistical forecasting model that belongs to the ARMA linear model class. According to \citep{hyndman2018forecasting}, the~development of exponential smoothing models depends on identifying the trend and seasonality present in the data. Conversely, ARIMA models are designed to characterize stationary, non-stationary, and~seasonal processes of order ($p$, $d$, $q$).

The utilization of ARIMA models for forecasting is often impeded by a prevalent barrier, namely the subjective nature and complexity of the order selection process \citep{hyndman2018forecasting}. Consequently, the~forecasting methodologies encompass the ARIMA procedure in the form of the following:
\begin{equation} \label{eq:arima}
(1 - \phi_{1}B)~(1 - \Phi_{1}B^{4}) (1 - B) (1 - B^{4})y_{t} = (1 + \theta_{1}B)~ (1 + \Theta_{1}B^{4})\varepsilon_{t} \text{,}
\end{equation}
where the variable $y_t$ denotes the true value at the $t$th instance, and $\varepsilon_t$ denotes the error sequence, which is considered a white noise process with a Gaussian distribution, specifically, the~error sequence has a mean of zero and a constant variance of $\sigma^2.$ The time series model is represented by the notation $\operatorname{ARIMA}(p, d, q)$. Identifying the model's appropriate order $(p, d, q)$ is critical to the ARIMA modeling~procedure.

ARIMA models can be fitted to both seasonal and non-seasonal data. Seasonal ARIMA demands a more detailed specification of the model's composition. Therefore, before~determining the estimation of the time series models, we performed the augmented Dickey--Fuller (ADF) \citep{dickey1979distribution} test, which could determine whether the dataset series are stationary; if the series is non-stationary,~data transformation is necessary. The~ADF statistics are obtained by $\Delta x_t = \alpha_0 + b_0 x_{t-1} + \sum\limits_{i=1}^k c_0 \Delta x_{t-1} + w_t$, where $\Delta$ is the difference operator; $\alpha_0$, $b_0$, and $c_0$ are coefficients to be estimated;~$x$ is the variable whose time series properties are examined, and $w$ is the white-noise error term. In~addition, the~null and the alternative hypotheses are, respectively, $b_0$ = 0 (series is non-stationary) and $b_0$ $<$ 0 (series is stationary). Table~\ref{tab:descriptive} provides an overview of the ADF test results, which indicates that the data are~stationary.

\subsubsection{Hybrid ETS-ANN~Model}
Hybrid methodologies, which combine two or more advanced techniques, have generally been observed to be more effective and enhance forecasting accuracy due to the inherent synergy from the individual methods \citep{hyndman2018forecasting}. However, the~efficacy of hybrid methodologies is contingent upon the effectiveness of their single models, which must mainly be chosen to assess the~problem. 

Historically, various statistical models, such as moving averages, exponential smoothing, and~autoregressive integrated moving averages (ARIMA) models, have been widely employed for time series forecasting. The~underlying assumption of these models is that the time series being analyzed originates from a linear process. However, the~nonlinearity of physical processes is a prevalent issue, as~\cite{zhang1998forecasting} noted. Therefore, researchers have developed several hybrid models to combine linear and nonlinear models, for~example, see~\cite{zhang2003time, babu2014moving, khandelwal2015time}. Accordingly, these models exhibited superior predictive performances compared to the individual models. To~this extent, in~their manuscript,~\cite{panigrahi2017hybrid} proposed a hybrid ETS-ANN model, which integrates linear and nonlinear exponential smoothing models derived from the innovation state space (ETS) with an artificial neural network (ANN) model. As~a result, they suggest a hybrid ETS-ANN model to identify various combinations of linear and nonlinear patterns in time series, such as:
\begin{equation} \label{eq1}
y_t=C_t^1+C_t^2
\end{equation}
\begin{equation}
e_t=y_t-\widehat{C_t^1}
\end{equation}
\begin{equation}
\hat{y}_t=\widehat{C_t^1}+\widehat{C_t^2}
\end{equation}

The model proposed by~\cite{panigrahi2017hybrid} suggests that the time series can be expressed as the sum of two components, denoted as $C_t^1$ and $C_t^2$ (as presented in Equation~(\ref{eq1})). The~components can exhibit either linear or nonlinear behavior, leading to three distinct categories of linear and nonlinear pattern~combinations.

\subsubsection{\textit{k}-Nearest Neighbor Classifier (\textit{k}NN)} 
The \textit{k}-nearest neighbors (\textit{k}-NN) method is a non-parametric machine learning technique that has been used for a considerable amount of time, rendering it a widely employed approach for classification \citep{fix1952discriminatory}. We are utilizing a substantial amount of training data, where a set of variables characterizes each data point. The~\textit{k}-NN algorithm assumes that proximate entities share similarities, specifically the K-nearest neighbors. This approach enables the algorithm to efficiently navigate the space and identify the most comparable items. The~algorithm's methodology is based on the square root of the N technique, which involves searching the entire volume of points in the training dataset. Given a specific point $x_0$ that needs to be classified into one of the K groups, a~possible approach is to determine the $k$ observed data points closest to $x_0$. According to~\cite{Neath2010-me}, the~sample with the most observed data points among the k-nearest neighbors is assigned $x_0$ in the classification form. Up~to this juncture, the~degree of similarity is contingent upon a specific distance metric, thereby rendering the efficacy of classifiers significantly reliant on the integrated distance metric \citep{weinberger2009distance}. The~methodology involves two distinct procedures, whereby the initial step entails the creation of an adjacency matrix and,~subsequently, the~estimation of edge weights is performed \citep{dornaika2017object}.

\subsubsection{Forecasting Performance~Measures}

We report on a number of robustness tests. In~addition to the mean absolute percentage error (MAPE), despite its widespread application in forecasting literature, we present forecasting results for the following loss functions: the mean absolute error (MAE) and the root mean square error (RMSE). The~results for each accuracy measure are reported in Table~\ref{tab:models}.
\begin{equation}
	\mathrm{MAE} = \frac{1}{h} \sum_{j=1}^h \left| y_{t+j} - \hat{y}_{t+j} \right|,
\end{equation}
\begin{equation}
	\mathrm{RMSE} = \sqrt{\frac{1}{h} \sum_{j=1}^h \left( y_{t+j} - \hat{y}_{t+j} \right)^2 },
\end{equation}
\begin{equation}
	\mathrm{MAPE} = \frac{1}{h} \sum_{j=1}^h \frac{y_{t+j} - \hat{y}_{t+j}}{y_{t+j}},
\end{equation}
where $\hat{y}_{t+j}$ refers to the model's forecast at time $t$. $y_{t+j}$ is the associated actual values, $h$ is the forecasting horizon, and~$j$ is the number of historical~observations. 

However, despite its widespread application in forecasting literature, the~mean absolute percentage error (MAPE) has attracted criticism as it exhibits significant shortcomings in generating undefined or infinite outcomes for actual zero or near-zero values \citep{kim2016new}. Therefore, to~assess the reviewed forecasting techniques and verify that the conclusions drawn from the study are indicative, we employed AvgRelMAE, considering the ARIMA model as a benchmark:
\begin{equation}
\operatorname{AvgRelMAE}=\exp \left(\frac{1}{N} \sum_{i=1}^N \ln \left(\frac{M A E_{\text {Model }, i}}{M A E_{\text {ARIMA}, i}}\right)\right)
\end{equation}
where $N$  represents the total number of series used for the evaluation of forecasting methods.

\section{Discussion}\label{sec4}

Table~\ref{tab:models} reports the accuracy metrics for three different time series forecasting models, namely,~ARIMA, ETS-ANN, and~\textit{k}NN, for~different SSMI financial market assets, i.e.,~BTC-USD, GDAXI, FTSE, N100, FCHI, and~SSMI, from~2018 to 2021. The~accuracy of the models is evaluated based on the following metrics:~mean absolute error (MAE), root mean square error (RMSE), and~the mean absolute percentage error (MAPE).

The findings indicate that the ranking of the models' performance remains consistent across each examined subperiod. Based on all accuracy metrics, it can be observed that the hybrid ETS-ANN model is the prevailing model for studying the specific dataset of financial market assets. According to the MAPE metric, the~ETS-ANN model had the lowest value for almost all indices and subperiods except for the entire period results. The \textit{k}NN model displayed some lower results during the pro-COVID-19 period of 2018 and 2019 for the BTC, FTSE, and~SSMI indices. In~contrast, ARIMA had a lower MAPE for almost all financial market assets for the entire period, except~the BTC market and the pro-COVID-19 period for the BTC, GDAXI, N100, and~FCHI indices. Therefore, the~choice of the most suitable model for a particular financial market asset and period depends on the specific characteristics of the data and the modeling assumptions. Overall, the~results suggest that the hybrid ETS-ANN model is a promising financial time series forecasting approach and can offer valuable perspectives to support investment~decision-making. 

More specifically, starting with BTC-USD, which represents the Bitcoin--USD exchange rate, we notice that all models show high error rates in predicting the exchange rate for all three years. Specifically, in~the 2018-–2021 forecast period, the~MAPE values for the three models ranged from 15.48\% to 28.45\%, indicating that the model's predictions were off by more than 15\% to 28\% of the actual value. The~best-performing model in this period is ETS-ANN, which achieves a MAPE of 15.48\%, indicating a moderate accuracy level. In~the year 2018, the~ETS-NN model had the highest MAE, RMSE, and~MAPE values, indicating less accurate forecasts compared to ARIMA and \textit{k}NN models. ARIMA performs better than both ETS-NN and \textit{k}NN in terms of every accuracy measure. In~2019, the~models' performance worsened, with~the MAPE values ranging from 8.58\% to 44.08\%. Similar to the previous period, the~best-performing model in 2019 was the ETS-ANN, which achieved a MAPE of 8.58\% and a low MAE comparable to other models, indicating a higher level of accuracy. In~2020, the~models' performance weakened, with~high MAPE values ranging from 22.51\% to 37.23\%, with~the \textit{k}NN showing the best performance. It is also worth noticing that during the first wave of the COVID outbreak in 2020, the~\textit{k}NN model achieved the lowest MAE, indicating a moderate level of accuracy. Similarly, in~2021, during~the ongoing pandemic, the~ETS-ANN model generated the best MAPE with a value of 8.39\%. Overall, the~results suggest that predicting BTC-USD is challenging for the selected models, and~the accuracy levels are generally low, especially in~2020.

Moving to GDAXI, we observe that the models' performance is relatively better than BTC-USD. For~the whole period, the~MAPE values for the three models ranged from 2.45\% to 9.52\%, indicating that the predictions were off by less than 10\% of the actual value. The~best-performing model is the ARIMA model, which achieves the lowest MAPE 2.45\% and the lowest MAE, indicating a good level of accuracy. Starting with the year 2018, the~\textit{k}NN and ARIMA models show similar forecast accuracies with lower MAE, RMSE, and~MAPE values compared to ETS-NN. In~2019, the~models' performance declined, with~the MAPE values ranging from 3.91\% to 5.89\%. The~best-performing model in 2019 was ARIMA, which achieved a MAPE of 3.91\%. In~2020, the~models' performance worsened even further, with~the MAPE values ranging from 1.35\% to 16.64\%. The best model, ETS-ANN, achieved the lowest MAPE, which was also similar in the post-COVID year 2021. Overall, the~results suggest that predicting GDAXI is more predictable than estimating the BTC-USD, and~the ETS-ANN and the ARIMA models perform in different subperiods to a great~extent.

The analysis shows that the ARIMA and ETS-ANN models perform better than the \textit{k}NN model for predicting the FTSE index's performance. As~such, the~models' performance is generally considered moderate, particularly during the periods of 2018--2021 and 2020. In~the entire period, the~MAPE values for the three models ranged from 2.38\% to 6.88\%, with~ARIMA being the best-performing model, followed by the ETS-ANN model (6.35\%). In~2018, the ETS-NN method demonstrated the best forecast accuracy with significantly low MAPE (1.70\%) compared to ARIMA and \textit{k}NN (3.59\%), respectively. In~2019, the~models' performance improved, with~the MAPE values ranging from 1.56\% to 2.00\%. In~2020, the~models' performance drastically dropped, with~the MAPE values ranging from 6.05\% to 8.18\%, indicating a considerable increase in prediction errors. The~best-performing model in 2020 was the ETS-ANN model, followed by \textit{k}NN, which achieved MAPE values of 6.05\% and 6.34\%, respectively. Similarly, in~2021, the~ETS-ANN outperformed the other~models.

As for the N100 index, the~models' performance is generally better than other indices. During~the 2018–-2021 period, MAPE values ranged from 4.40\% to 12.37\%, with~ARIMA outperforming the other models. In~2019, the~models' performance was better, with~the MAPE values ranging from 2.91\% to 4.00\%. In~2018, both ARIMA and ETS-NN methods showed quite similar forecast accuracies with lower MAPE values than \textit{k}NN, with~ETS-NN (MAPE: 3.76\%) outperforming the other models. Similarly, the~best-performing model in 2019 was ARIMA, which achieved a MAPE of 2.91\%. In~2020, the~models' performance dropped due to the pandemic, with~the MAPE values ranging from 7.71\% to 10.54\%, with~\textit{k}NN outperforming the other models, followed closely by the ETS-ANN model. In~the year 2021, ETS-ANN showed more robust accuracy results, surpassing the other two models with a MAPE of 2.94\%, and demonstrating better performance across all other measures compared to the other models.
The FCHI index indicates that the models' performance is generally weak, especially during the pandemic year 2020. The~analysis shows that the ARIMA and ETS-ANN models performed better over more periods than the \textit{k}NN model. For~the 2018--2021 period, ARIMA generated the most robust predictive results, outperforming the other models, with a MAPE value of 5.24\%, while ETS-ANN provided the worst accuracy (11.27\%). In~2018, both ARIMA and ETS-NN models exhibited similar forecast accuracies, with slightly lower MAE and RMSE values compared to \textit{k}NN, with~ETS-NN outperforming ARIMA. In~2019, the~models' performance slightly improved, with~the MAPE values ranging from 3.56\% to 4.72\%, with~a similar performance ranking ARIMA as the best model, followed by the hybrid ETS-ANN model. In~2020, the~models' performance drastically dropped, with~the MAPE values ranging from 8.18\% to 9.66\%, indicating a considerable increase in prediction errors. The~best-performing model in 2020 was the hybrid ETS-ANN model, followed by \textit{k}NN, which achieved a MAPE value of 9.01\%. In~the end, the~ETS-ANN outperformed the other models in the year~2021.

Finally, the~SSMI index performed worse than the comparable indices, except~the BTC, for~almost every period. The~models' performance was generally weak, especially in 2018--2021 and post-pandemic 2021. In~the 2018--2021 period, the~MAPE values for the three models ranged from 7.03\% to 8.54\%, with~ARIMA being the best-performing model. However, the~\textit{k}NN model generated almost similar MAE accuracy to ARIMA for the same period. Starting in the year 2018, all three models showed comparable forecast accuracies for the SSMI index, with~slightly lower MAE and RMSE values for \textit{k}NN, and ARIMA showing equal MAPE with the ETS-NN (1.81\%). In~2019, the~models' performance slightly improved, with~the \textit{k}NN model generating a more robust MAPE value of 3.73\% and similarly better MAE performance. During~the pandemic, the~models' performance surprisingly improved, with~MAPE values ranging from 0.97\% to 6.22\%, indicating a considerable and robust increase in prediction errors. The~best-performing model in 2020 was the ETS-ANN model, followed by the \textit{k}NN model, which achieved a MAPE of 2.98\%. Finally, the~ETS-ANN outperformed the other models in 2021 with more robust MAPE and MAE accuracies. The~analysis shows that the ARIMA and \textit{k}NN models performed with high precision, similar to the ETS-ANN model, except~for one~period.

Following the previous analysis of the financial market assets, we evaluated the performances of three different time series forecasting models, namely, the~ARIMA, ETS-ANN, and~\textit{k}NN, from~2018 to 2021. The~findings indicate that the hybrid ETS-ANN model is the prevailing choice for examining the specific dataset of financial market assets, with~the lowest MAPE for almost all indices and subperiods except for the 2018--2021 period results. On~the other side, MAPE received criticism due to some drawbacks; thus, considering MAE, we observe that the ARIMA and \textit{k}NN models demonstrated similar performance in several subperiods. Therefore, the~choice of the most suitable model depends on the specific characteristics of the data and the modeling assumptions. In~conclusion, the~hybrid ETS-ANN model is a promising financial time series forecasting approach that can provide valuable investment decision~insights.

\begin{table}[H]
\caption{{Forecasting model performances~} 
 for the financial markets of each~period.}
\label{tab:models}
\renewcommand{\arraystretch}{1}
\setlength{\tabcolsep}{6pt}
\resizebox{\textwidth}{!}{%
\begin{tabular}{lrrrrrrrrr}
\toprule
         & \multicolumn{3}{c}{\textbf{ARIMA}}                                                     & \multicolumn{3}{c}{\textbf{ETS-NN}}                                                    & \multicolumn{3}{c}{\textbf{\textit{k}NN}}                                                       \\ \cmidrule{2-10}
         & \multicolumn{1}{c}{\textbf{MAE}} & \multicolumn{1}{c}{\textbf{RMSE}} & \multicolumn{1}{c}{\textbf{MAPE}} & \multicolumn{1}{c}{\textbf{MAE}} & \multicolumn{1}{c}{\textbf{RMSE}} & \multicolumn{1}{c}{\textbf{MAPE}} & \multicolumn{1}{c}{\textbf{MAE}} & \multicolumn{1}{c}{\textbf{RMSE}} & \multicolumn{1}{c}{\textbf{MAPE}} \\ \midrule
\multicolumn{2}{l}{Year 2018--2021} &                          &                          &                         &                          &                          &                         &                          &                          \\
BTC-USD  & 13,471.98                & 15,702.97                 & 25.44                    & 6701.90                 & 7862.06                  & 15.48                    & 22,518.47                & 23,801.17                 & 28.45                    \\
GDAXI    & 386.32                  & 458.67                   & 2.45                     & 1290.97                 & 1319.07                  & 8.27                     & 1253.96                 & 1639.35                  & 9.52                     \\
FTSE     & 171.47                  & 205.32                   & 2.38                     & 453.42                  & 469.57                   & 6.35                     & 461.69                  & 564.05                   & 6.88                     \\
N100     & 58.16                   & 69.84                    & 4.40                     & 121.69                  & 127.14                   & 9.34                     & 127.21                  & 147.98                   & 12.37                    \\
FCHI     & 357.04                  & 425.17                   & 5.24                     & 755.48                  & 786.81                   & 11.27                    & 503.26                  & 646.80                   & 9.15                     \\
SSMI     & 859.91                  & 983.97                   & 7.03                     & 944.07                  & 1022.29                  & 7.78                     & 884.45                  & 1110.99                  & 8.54                     \\ \midrule
\multicolumn{2}{l}{Year 2018}      &                          &                          &                         &                          &                          &                         &                          &                          \\
BTC-USD  & 893.39                  & 1297.50                  & 19.42                    & 2906.60                 & 3030.60                  & 72.02                    & 1983.54                 & 2480.08                  & 37.19                    \\
GDAXI    & 529.96                  & 689.04                   & 4.54                     & 716.69                  & 753.22                   & 6.51                     & 522.10                  & 682.16                   & 4.53                     \\
FTSE     & 261.75                  & 309.93                   & 3.59                     & 115.51                  & 156.82                   & 1.70                     & 255.81                  & 300.29                   & 3.59                     \\
N100     & 37.81                   & 47.12                    & 3.78                     & 35.38                   & 41.67                    & 3.76                     & 44.13                   & 52.73                    & 4.47                     \\
FCHI     & 188.01                  & 233.41                   & 3.65                     & 182.92                  & 212.41                   & 3.78                     & 232.37                  & 294.40                   & 4.61                     \\
SSMI     & 158.89                  & 202.98                   & 1.81                     & 158.99                  & 187.17                   & 1.81                     & 172.57                  & 204.05                   & 1.94                     \\ \midrule
\multicolumn{2}{l}{Year 2019}      &                          &                          &                         &                          &                          &                         &                          &                          \\
BTC-USD  & 1468.02                 & 1659.53                  & 17.50                    & 644.74                  & 717.56                   & 8.58                     & 3895.68                 & 4603.57                  & 44.08                    \\
GDAXI    & 459.71                  & 605.76                   & 3.61                     & 777.87                  & 784.44                   & 5.89                     & 594.54                  & 701.98                   & 4.76                     \\
FTSE     & 123.80                  & 148.80                   & 1.70                     & 138.75                  & 194.79                   & 1.85                     & 113.94                  & 141.66                   & 1.56                     \\
N100     & 32.24                   & 39.62                    & 2.93                     & 38.34                   & 40.08                    & 3.38                     & 42.75                   & 48.54                    & 4.00                     \\
FCHI     & 202.94                  & 255.31                   & 3.55                     & 269.60                  & 277.26                   & 4.55                     & 267.13                  & 320.31                   & 4.72                     \\
SSMI     & 344.25                  & 438.14                   & 3.38                     & 493.55                  & 509.50                   & 4.72                     & 371.20                  & 423.61                   & 3.73                     \\ \midrule
\multicolumn{2}{l}{Year 2020}      &                          &                          &                         &                          &                          &                         &                          &                          \\
BTC-USD  & 4412.44                 & 6286.24                  & 26.43                    & 7256.93                 & 7863.49                  & 37.23                    & 3067.44                 & 4958.43                  & 22.51                    \\
GDAXI    & 818.08                  & 909.35                   & 6.42                     & 175.36                  & 257.79                   & 1.35                     & 723.74                  & 1137.54                  & 6.42                     \\
FTSE     & 230.45                  & 303.38                   & 3.86                     & 389.48                  & 411.95                   & 6.05                     & 399.32                  & 517.23                   & 6.34                     \\
N100     & 35.59                   & 43.82                    & 3.55                     & 86.94                   & 90.54                    & 7.97                     & 78.33                   & 97.47                    & 7.71                     \\
FCHI     & 254.47                  & 291.15                   & 5.20                     & 450.81                  & 472.97                   & 8.18                     & 455.62                  & 529.93                   & 9.01                     \\
SSMI     & 200.22                  & 239.08                   & 1.96                     & 101.48                  & 131.54                   & 0.97                     & 288.88                  & 507.73                   & 2.98                     \\ \midrule
\multicolumn{2}{l}{Year 2021}      &                          &                          &                         &                          &                          &                         &                          &                          \\
BTC-USD  & 6944.45                 & 8983.94                  & 13.41                    & 4425.80                 & 5468.44                  & 8.39                     & 7181.29                 & 9254.76                  & 17.71                    \\
GDAXI    & 389.88                  & 462.52                   & 2.48                     & 284.25                  & 344.78                   & 1.81                     & 611.14                  & 911.19                   & 4.24                     \\
FTSE     & 199.91                  & 230.71                   & 2.78                     & 166.57                  & 186.08                   & 2.28                     & 227.05                  & 259.67                   & 3.24                     \\
N100     & 59.84                   & 71.61                    & 4.53                     & 39.93                   & 45.18                    & 2.94                     & 74.28                   & 93.97                    & 6.14                     \\
FCHI     & 336.90                  & 405.34                   & 4.94                     & 336.55                  & 366.36                   & 4.78                     & 426.25                  & 540.26                   & 6.91                     \\
SSMI     & 861.34                  & 985.47                   & 7.04                     & 446.78                  & 497.18                   & 3.55                     & 527.14                  & 725.35                   & 4.69      \\ \bottomrule              
\end{tabular}%
}
\noindent{\footnotesize{Note: Bitcoin (BTC-USD), German DAX (GDAXI), Financial Times Stock Exchange (FTSE), Euronext (N100), French Stock Market (FCHI), Swiss SSMI (SSMI), mean absolute error (MAE), root-mean-square error (RMSE), mean absolute percentage error (MAPE).}}
\end{table}

Table~\ref{tab:AvgRelMAE} shows the average relative MAE results per year. The~results reveal that the ETS-NN and \textit{k}NN models outperformed ARIMA in most cases. Therefore, ETS-NN had the lowest AvgRelMAE for all asset indices in 2020 during the outbreak and in 2021, while \textit{k}NN exceeded ARIMA and had a low MAE in 2020. However, ARIMA outperformed both models during 2018--2021, except~the ETS-NN model for BTC, and~generated the lowest AvgRelMAE for almost every asset index, except BTC and FTSE in 2019. In~2018, we observed that ARIMA outperformed, in~many cases, the~other methods, except~the ETS-NN model for the FTSE, N100, and~FCHI indices. Similarly, \textit{k}NN generated better forecasting accuracy than ARIMA in the GDAXI and FTSE indices. Finally, ARIMA outperformed the \textit{k}NN model for every asset index except the BTC and SSMI in 2021. Once again, the~performances of the models varied across financial market indices and years, highlighting the importance of selecting the appropriate model for a particular market and~period.

\begin{table}[H]
\caption{Average relative MAE performance of the forecasting~methods.}
\label{tab:AvgRelMAE}
\renewcommand{\arraystretch}{1}
\setlength{\tabcolsep}{17pt}
\resizebox{\textwidth}{!}{%
\begin{tabular}{lrrrrrr}
\toprule
\textbf{Model}  & \multicolumn{1}{c}{\textbf{BTC-USD}} & \multicolumn{1}{c}{\textbf{GDAXI}} & \multicolumn{1}{c}{\textbf{FTSE}} & \multicolumn{1}{c}{\textbf{N100}} & \multicolumn{1}{c}{\textbf{FCHI}} & \multicolumn{1}{c}{\textbf{SSMI}} \\ \midrule
\multicolumn{2}{l}{Year 2018--2021}   &                           &                          &                          &                          &                          \\
ETS-NN & 0.497                       & 3.342                     & 2.644                    & 2.092                    & 2.116                    & 1.098                    \\
\textit{k}NN    & 1.672                       & 3.246                     & 2.693                    & 2.187                    & 1.410                    & 1.029                    \\
ARIMA  & 1.000                       & 1.000                     & 1.000                    & 1.000                    & 1.000                    & 1.000                    \\ \midrule
\multicolumn{2}{l}{Year 2018}        &                           &                          &                          &                          &                          \\
ETS-NN & 3.253                       & 1.352                     & 0.441                    & 0.936                    & 0.973                    & 1.001                    \\
\textit{k}NN    & 2.220                       & 0.985                     & 0.977                    & 1.167                    & 1.236                    & 1.086                    \\
ARIMA  & 1.000                       & 1.000                     & 1.000                    & 1.000                    & 1.000                    & 1.000                    \\ \midrule
\multicolumn{2}{l}{Year 2019}        &                           &                          &                          &                          &                          \\
ETS-NN & 0.063                       & 1.557                     & 0.955                    & 1.198                    & 1.325                    & 1.252                    \\
\textit{k}NN    & 0.382                       & 1.190                     & 0.784                    & 1.336                    & 1.313                    & 0.942                    \\
ARIMA  & 1.000                       & 1.000                     & 1.000                    & 1.000                    & 1.000                    & 1.000                    \\ \midrule
\multicolumn{2}{l}{Year 2020}        &                           &                          &                          &                          &                          \\
ETS-NN & 1.634                       & 0.082                     & 0.768                    & 0.807                    & 0.904                    & 0.159                    \\
\textit{k}NN    & 0.690                       & 0.339                     & 0.788                    & 0.727                    & 0.913                    & 0.452                    \\
ARIMA  & 1.000                       & 1.000                     & 1.000                    & 1.000                    & 1.000                    & 1.000                    \\ \midrule
\multicolumn{2}{l}{Year 2021}        &                           &                          &                          &                          &                          \\
ETS-NN & 0.321                       & 0.729                     & 0.851                    & 0.644                    & 0.880                    & 0.511                    \\
\textit{k}NN    & 0.521                       & 1.566                     & 1.160                    & 1.197                    & 1.115                    & 0.603                    \\
ARIMA  & 1.000                       & 1.000                     & 1.000                    & 1.000                    & 1.000                    & 1.000  \\ \bottomrule                 
\end{tabular}%
}
\end{table}

\section{Conclusions}
According to the literature, studies evaluating the European financial markets' performance are scant because of the extensive number of publications focusing on the U.S. market. This study aims to compare parametric and non-parametric forecasting models, such as the ARIMA, Hybrid ETS-ANN, and~\textit{k}NN predictive models, in~predicting the key European stock market indices and comparing them to the cryptocurrency market for an extended period, including the pandemic in 2020. The~study period was divided into three subperiods: the pre-pandemic period, the~pandemic period, and~the continuous~period.

It is worth noting that the performances of the models varied across different indices and years, indicating that the choice of the best model should depend on the specific context and data. Overall, the~ARIMA and ETS-ANN models performed better than the \textit{k}NN model while evaluating the model's accuracy. For~instance, the~ARIMA model ranks better and unexpectedly outperforms the hybrid ETS-ANN and \textit{k}NN machine learning models. The~ARIMA model performed better for the BTC and GDAXI indices before COVID-19, as~it is a commonly used model in finance and has been shown to perform well in time series forecasting. The~cryptocurrency market is known for its fluctuation and unpredictability, making it challenging to forecast~accurately.

Unsurprisingly, ARIMA performed well in the years 2018 and 2019. On~the other hand, the~\textit{k}NN model showed better results for the FTSE and SSMI indices during the pro-COVID-19 period. It is also interesting to see that \textit{k}NN outperformed the other models in 2020 because it is not a traditional time series forecasting model but rather a machine learning algorithm for classification and~regression.

In conclusion, the~results suggest that the ETS-ANN model is the best-performing model despite its moderate level of accuracy. The~low MAPE values across almost all subperiods indicate that the model predictions perform well in different subperiods. Therefore, it is essential to carefully evaluate the performances of other models and choose the one that provides the most reliable and accurate forecast for the specific~application.

As with any study, there are limitations. For~example, although~we used different hyperparameters for the ARIMA and \textit{k}NN models based on the year and financial index in this study, we used the same hyperparameters across years for the hybrid ETS-ANN model. Further research could investigate the performances of other forecasting models or consider additional variables and parameters to improve the accuracies of financial market predictions. In~addition, researchers could analyze how past events such as COVID-19 have impacted the financial sector and compare financial indices to a broader range of cryptocurrencies. Comparing COVID-19 to past events could create a portfolio suitable for today's highly interconnected and unpredictable economic~landscape.

\vspace{9pt}

\funding{This research received no external~funding.}

\dataavailability
{Data are publicly available at \url{https://www.cryptodatadownload.com/data/}}

\conflictsofinterest{The author declares no conflict of~interest.}


\abbreviations{Abbreviations}{
The following abbreviations are used in this manuscript:\\

\noindent 
\begin{tabular}{@{}ll}
ACF & autocorrelation function \\
ADF & augmented Dickey--Fuller\\
AIC & Akaike information criterion \\
ANN & artificial neural network \\
ARCH & autoregressive conditional heteroskedasticity\\
ARIMA & autoregressive integrated moving average\\
AvgRelMAE & average relative mean absolute error \\
ETS & exponential smoothing \\
h & horizon \\
kNN & k-nearest neighbor classifier \\
MAE & mean absolute error \\
MAPE & mean absolute percentage error \\
RMSE & root mean square error \\
\end{tabular}}

\appendix
\section*{Appendix A. Forecasting the Model Parameters}

Table~\ref{tab:parameters} provides the values of the parameters of each~model.

\setcounter{table}{0}
\renewcommand{\thetable}{A\arabic{table}}


\begin{table}[H]
\caption{Specifications of the forecasting~models.}
\setlength{\tabcolsep}{13pt}
\label{tab:parameters}
\resizebox{\textwidth}{!}{%
\begin{tabular}{lcrrrrr}
\toprule
        &               & \multicolumn{1}{l}{\textbf{Year}}        & \multicolumn{1}{l}{}       & \multicolumn{1}{l}{}       & \multicolumn{1}{l}{}      & \multicolumn{1}{l}{}       \\
\textbf{Asset}   & \textbf{Model}         & \multicolumn{1}{c}{\textbf{2018--2021}} & \multicolumn{1}{c}{\textbf{2018}}   & \multicolumn{1}{c}{\textbf{2019}}   & \multicolumn{1}{c}{\textbf{2020}}  & \multicolumn{1}{c}{\textbf{2021}}   \\ \midrule
BTC-USD & ARIMA         & (4, 1, 1)                       & (2, 1, 2)                  & (1, 0, 5)                  & (4, 1, 4)                 & (3, 0, 3)                  \\
GDAXI   &               & (1, 1, 1)                       & (3, 1, 2)                  & (1, 1, 5)                  & (3, 0, 4)                 & (1, 1, 4)                  \\
FTSE    &               & (4, 1, 3)                       & (5, 0, 2)                  & (5, 1, 1)                  & (3, 0, 5)                 & (2, 1, 4)                  \\
N100    &               & (5, 1, 0)                       & (3, 1, 2)                  & (2, 1, 1)                  & (1, 0, 0)                 & (1, 1, 5)                  \\
FCHI    &               & (5, 1, 4)                       & (3, 0, 4)                  & (2, 1, 2)                  & (3, 0, 2)                 & (1, 1, 4)                  \\
SSMI    &               & (1, 1, 5)                       & (2, 1, 5)                  & (5, 1, 0)                  & (4, 0, 2)                 & (4, 1, 5)                  \\
\multicolumn{6}{l}{{Note}
: optimal parameters based on information criterion `AIC'.} & \\\midrule
BTC-USD & kNN           & k = 26      & k = 27 & k = 26 & k = 9 & k = 27 \\
GDAXI   &               & k = 19                          & k = 27                     & k = 18                     & k = 21                    & k = 18                     \\
FTSE    &               & k = 25                          & k = 27                     & k = 27                     & k = 16                    & k = 25                     \\
N100    &               & k = 27                          & k = 20                     & k = 23                     & k = 27                    & k = 21                     \\
FCHI    &               & k = 14                          & k = 24                     & k = 20                     & k = 6                     & k = 18                     \\
SSMI    &               & k = 27                          & k = 22                     & k = 15                     & k = 14                    & k = 27                     \\
\multicolumn{6}{l}{{Note}
: weights: uniform; algorithm: brute; p: 2; k: optimal k.}       &                            \\\midrule
ETS-NN  & ETS           & \multicolumn{1}{c}{NN}          &                            &                            &                           &                            \\
        & ETS(A, Ad, N) & \multicolumn{2}{l}{LSTM layers: 50}                           &                            &                           &                            \\
        &               & \multicolumn{2}{l}{Dropout rate: 0.2}                        &                            &                           &                            \\
        &               & \multicolumn{2}{l}{Optimizer: Adam}                          &                            &                           &                            \\
        &               & \multicolumn{3}{l}{Loss Function: Mean Squared Error}                                     &                           &                            \\
        &               & \multicolumn{2}{l}{Number of Epochs: 100}                    &                            &                           &                            \\
        &               & \multicolumn{2}{l}{Batch Size: 32}                           &                            &          &   \\ \bottomrule                        
\end{tabular}%
}
\end{table}

\begin{adjustwidth}{-\extralength}{0cm}

\reftitle{References}

\begin{thebibliography}{999}

\bibitem[Al-Awadhi \em{et~al.}(2020)Al-Awadhi, Abdullah M.,  Alsaifi, Khaled, Al-Awadhi, Ahmad, and Alhammadi, Salah]{al2020death}
Al-Awadhi, A. M.;  Alsaifi, K.; Al-Awadhi, A.; Alhammadi, S.
\newblock Death and contagious infectious diseases: Impact of the COVID-19 virus on stock market returns.
\newblock {\em Journal of behavioral and experimental finance} {\bf 2020}, {\em 27},~100326.

\bibitem[Ampountolas(2022)]{ijfs10030051}
Ampountolas, A.
\newblock Cryptocurrencies Intraday High-Frequency Volatility Spillover Effects Using Univariate and Multivariate GARCH Models.
\newblock {\em International Journal of Financial Studies} {\bf 2022}, {\em
  51}, {2227--7072.}

\bibitem[Stolberg(2022)]{NYT2022COVID}
Stolberg, S.G.
\newblock Biden Says the Pandemic Is Over. But at Least 400 People Are Dying
  Daily.   \emph{The New York Times}, {19 September 2022}.
\newblock
Available online: \url{https://www.nytimes.com/2022/09/19/us/politics/biden-covid-pandemic-over.html}  (accessed on 28 September 2022).  

\bibitem[Fisher \em{et~al.}(2021)Fisher, Roberts, McKinlay, Fancourt, and
  Burton]{fisher2021impact}
Fisher, A.; Roberts, A.; McKinlay, A.R.; Fancourt, D.; Burton, A.
\newblock The impact of the COVID-19 pandemic on mental health and well-being
  of people living with a long-term physical health condition: A qualitative
  study.
\newblock {\em BMC Public Health} {\bf 2021}, {\em 21},~1801.

\bibitem[Ashraf(2020)]{ashraf2020economic}
Ashraf, B.N.
\newblock Economic impact of government interventions during the COVID-19
  pandemic: International evidence from financial markets.
\newblock {\em J. Behav. Exp. Financ.} {\bf 2020}, {\em
  27},~100371.

\bibitem[Watson and Popescu(2021)]{watson2021will}
Watson, R.; Popescu, G.H.
\newblock Will the COVID-19 pandemic lead to long-term consumer perceptions, behavioral intentions, and acquisition decisions?
\newblock {\em Econ. Manag. Financ. Mark.} {\bf 2021}, {\em
  16}, {70--83.} 


\bibitem[Mazur \em{et~al.}(2021)Mazur, Dang, and Vega]{MAZUR2021101690}
Mazur, M.; Dang, M.; Vega, M.
\newblock COVID-19 and the march 2020 stock market crash. Evidence from
  S\&P500.
\newblock {\em Financ. Res. Lett.} {\bf 2021}, {\em 38},~101690.
\newblock {\url{https://doi.org/https://doi.org/10.1016/j.frl.2020.101690}}.

\bibitem[Goldstein \em{et~al.}(2021)Goldstein, Koijen, and
  Mueller]{goldstein2021covid}
Goldstein, I.; Koijen, R.S.; Mueller, H.M.
\newblock COVID-19 and its impact on financial markets and the real economy.
\newblock {\em  Rev. Financ. Stud.} {\bf 2021}, {\em
  34},~5135--5148.

\bibitem[Di and Xu(2022)]{di2022covid}
Di, M.; Xu, K.
\newblock COVID-19 vaccine and post-pandemic recovery: Evidence from Bitcoin
  cross-asset implied volatility spillover.
\newblock {\em Financ. Res. Lett.} {\bf 2022}, {\em 50},~103289.

\bibitem[Aslam \em{et~al.}(2020)Aslam, Mohti, and Ferreira]{aslam2020evidence}
Aslam, F.; Mohti, W.; Ferreira, P.
\newblock Evidence of intraday multifractality in European stock markets during
  the recent coronavirus (COVID-19) outbreak.
\newblock {\em Int. J. Financ. Stud.} {\bf 2020}, {\em
  8},~31.

\bibitem[Khattak \em{et~al.}(2021)Khattak, Ali, and
  Rizvi]{khattak2021predicting}
Khattak, M.A.; Ali, M.; Rizvi, S.A.R.
\newblock Predicting the European stock market during COVID-19: A machine
  learning approach.
\newblock {\em MethodsX} {\bf 2021}, {\em 8},~101198.

\bibitem[Su \em{et~al.}(2022)Su, Rizvi, Naqvi, Mirza, and Umar]{su2022covid19}
Su, C.W.; Rizvi, S.K.A.; Naqvi, B.; Mirza, N.; Umar, M.
\newblock COVID19: A blessing in disguise for European stock markets?
\newblock {\em Financ. Res. Lett.} {\bf 2022}, {\em 49},~103135.

\bibitem[Lachaab and Omri(2023)]{lachaab2023machine}
Lachaab, M.; Omri, A.
\newblock Machine and deep learning-based stock price prediction during the COVID-19 pandemic: The case of CAC 40 index.
\newblock {\em EuroMed J. Bus.} {\bf 2023}, \emph{{ahead-of-print}
}. \url{https://doi.org/10.1108/EMJB-05-2022-0104}.

\bibitem[Topcu and Gulal(2020)]{topcu2020impact}
Topcu, M.; Gulal, O.S.
\newblock The impact of COVID-19 on emerging stock markets.
\newblock {\em Financ. Res. Lett.} {\bf 2020}, {\em 36},~101691.

\bibitem[Azimli(2020)]{azimli2020impact}
Azimli, A.
\newblock The impact of COVID-19 on the degree of dependence and structure of
  risk-return relationship: A quantile regression approach.
\newblock {\em Financ. Res. Lett.} {\bf 2020}, {\em 36},~101648.

\bibitem[Goodell and Goutte(2021)]{goodell2021co}
Goodell, J.W.; Goutte, S.
\newblock Co-movement of COVID-19 and Bitcoin: Evidence from wavelet coherence
  analysis.
\newblock {\em Financ. Res. Lett.} {\bf 2021}, {\em 38},~101625.

\bibitem[Akhtaruzzaman \em{et~al.}(2022)Akhtaruzzaman, Boubaker, Nguyen, and
  Rahman]{akhtaruzzaman2022systemic}
Akhtaruzzaman, M.; Boubaker, S.; Nguyen, D.K.; Rahman, M.R.
\newblock Systemic risk-sharing framework of cryptocurrencies in the COVID-19  crisis.
\newblock {\em Financ. Res. Lett.} {\bf 2022},  \emph{47},  102787.

\bibitem[Uddin \em{et~al.}(2022)Uddin, Yahya, Goswami, Lucey, and
  Ahmed]{uddin2022stock}
Uddin, G.S.; Yahya, M.; Goswami, G.G.; Lucey, B.; Ahmed, A.
\newblock Stock market contagion during the COVID-19 pandemic in emerging
  economies.
\newblock {\em Int. Rev. Econ. Financ.} {\bf 2022}, {\em
  79},~302--309.

\bibitem[Ampountolas(2023)]{ampountolas2023effect}
Ampountolas, A.
\newblock The Effect of COVID-19 on Cryptocurrencies and the Stock Market
  Volatility: A Two-Stage DCC-EGARCH Model Analysis.
\newblock {\em J. Risk Financ. Manag.} {\bf 2023}, {\em
  16},~25.

\bibitem[Hyndman and Khandakar(2008)]{hyndman2008automatic}
Hyndman, R.J.; Khandakar, Y.
\newblock Automatic time series forecasting: The forecast package for R.
\newblock {\em J. Stat. Softw.} {\bf 2008}, {\em 27},~1--22.

\bibitem[Panigrahi and Behera(2017)]{panigrahi2017hybrid}
Panigrahi, S.; Behera, H.S.
\newblock A hybrid ETS--ANN model for time series forecasting.
\newblock {\em Eng. Appl. Artif. Intell.} {\bf 2017},
  {\em 66},~49--59.

\bibitem[Leippold \em{et~al.}(2022)Leippold, Wang, and
  Zhou]{leippold2022machine}
Leippold, M.; Wang, Q.; Zhou, W.
\newblock Machine learning in the Chinese stock market.
\newblock {\em J. Financ. Econ.} {\bf 2022}, {\em 145},~64--82.


\bibitem[Ciner(2021)]{CINER2021101705}
Ciner, C.
\newblock Stock return predictability in the time of COVID-19.
\newblock {\em Financ. Res. Lett.} {\bf 2021}, {\em 38},~101705.
\newblock {\url{https://doi.org/https://doi.org/10.1016/j.frl.2020.101705}}.

\bibitem[Liu \em{et~al.}(2020)Liu, Manzoor, Wang, Zhang, and
  Manzoor]{liu2020covid}
Liu, H.; Manzoor, A.; Wang, C.; Zhang, L.; Manzoor, Z.
\newblock The COVID-19 outbreak and affected countries stock markets response.
\newblock {\em Int. J. Environ. Res. Public Health} {\bf 2020}, {\em 17},~2800.

\bibitem[Barker(2020)]{barker2020machine}
Barker, J.
\newblock Machine learning in M4: What makes a good unstructured model?
\newblock {\em Int. J. Forecast.} {\bf 2020}, {\em
  36},~150--155.

\bibitem[Shanker \em{et~al.}(1996)Shanker, Hu, and Hung]{shanker1996effect}
Shanker, M.; Hu, M.Y.; Hung, M.S.
\newblock Effect of data standardization on neural network training.
\newblock {\em Omega} {\bf 1996}, {\em 24},~385--397.

\bibitem[Shankar \em{et~al.}(2016)Shankar, Singh, and
  Howard]{shankar2016neural}
Shankar, K.H.; Singh, I.; Howard, M.W.
\newblock Neural mechanism to simulate a scale-invariant future.
\newblock {\em Neural Comput.} {\bf 2016}, {\em 28}, 2594--2627.





\bibitem[Hyndman and Athanasopoulos(2018)]{hyndman2018forecasting}
Hyndman, R.J.; Athanasopoulos, G.
\newblock {\em Forecasting: Principles and Practice}; {OTexts: Melbourne, Australia. OTexts.com/fpp2}
:  2018.

\bibitem[Dickey and Fuller(1979)]{dickey1979distribution}
Dickey, D.A.; Fuller, W.A.
\newblock Distribution of the estimators for autoregressive time series with a
  unit root.
\newblock {\em J. Am. Stat. Assoc.} {\bf 1979},
  {\em 74},~427--431.

\bibitem[Zhang \em{et~al.}(1998)Zhang, Patuwo, and Hu]{zhang1998forecasting}
Zhang, G.; Patuwo, B.E.; Hu, M.Y.
\newblock Forecasting with artificial neural networks:: The state of the art.
\newblock {\em Int. J. Forecast.} {\bf 1998}, {\em
  14},~35--62.

\bibitem[Zhang(2003)]{zhang2003time}
Zhang, G.P.
\newblock Time series forecasting using a hybrid ARIMA and neural network
  model.
\newblock {\em Neurocomputing} {\bf 2003}, {\em 50},~159--175.

\bibitem[Babu and Reddy(2014)]{babu2014moving}
Babu, C.N.; Reddy, B.E.
\newblock A moving-average filter based hybrid ARIMA--ANN model for forecasting
  time series data.
\newblock {\em Appl. Soft Comput.} {\bf 2014}, {\em 23},~27--38.

\bibitem[Khandelwal \em{et~al.}(2015)Khandelwal, Adhikari, and
  Verma]{khandelwal2015time}
Khandelwal, I.; Adhikari, R.; Verma, G.
\newblock Time series forecasting using hybrid ARIMA and ANN models based on
  DWT decomposition.
\newblock {\em Procedia Comput. Sci.} {\bf 2015}, {\em 48},~173--179.

\bibitem[Fix and Hodges~Jr(1952)]{fix1952discriminatory}
Fix, E.; Hodges~Jr, J.L.
\newblock \emph{Discriminatory Analysis-Nonparametric Discrimination: Small Sample
  Performance};
\newblock Technical Report; {University of California, Berkeley}
: {Berkeley, CA, USA}
,  1952.

\bibitem[Neath \em{et~al.}(2010)Neath, Johnson, Baker, McGaw, and
  Peterson]{Neath2010-me}
Neath, R.C.; Johnson, M.S.; Baker, E.; McGaw, B.; Peterson, P.
\newblock Discrimination and Classification. In {\em International Encyclopedia
  of Education}, 3rd ed.;  {Baker, E., McGaw, B., Peterson, P.}, Eds.; Elsevier
  Ltd.: London, UK,  2010; Volume~1, pp. 135--141.

\bibitem[Weinberger and Saul(2009)]{weinberger2009distance}
Weinberger, K.Q.; Saul, L.K.
\newblock Distance metric learning for large margin nearest neighbor classification.
\newblock {\em J. Mach. Learn. Res.} {\bf 2009}, {\em 10}, {207--244}.


\bibitem[Dornaika \em{et~al.}(2017)Dornaika, Bosaghzadeh, Salmane, and
  Ruichek]{dornaika2017object}
Dornaika, F.; Bosaghzadeh, A.; Salmane, H.; Ruichek, Y.
\newblock Object Categorization Using Adaptive Graph-Based Semi-supervised
  Learning. In {\em Handbook of Neural Computation}; Elsevier: {Amsterdam, The Netherlands,} 
  2017; pp.  167--179.

\bibitem[Kim and Kim(2016)]{kim2016new}
Kim, S.; Kim, H.
\newblock A new metric of absolute percentage error for intermittent demand
  forecasts.
\newblock {\em Int. J. Forecast.} {\bf 2016}, {\em
  32},~669--679.



\end{thebibliography}

\PublishersNote{}
\end{adjustwidth}
\end{document}